\documentclass[review]{elsarticle}

\usepackage[english]{babel}

\usepackage{amssymb}
\usepackage{amsmath}
\usepackage{amsfonts}
\usepackage{graphicx}
\usepackage[colorlinks=true, allcolors=blue]{hyperref}

\journal{Materials Characterization}

\begin{document}
\begin{frontmatter}
\title{NLSTEM: Non-local denoising for enhanced 4D-STEM pattern indexing}
\author[1]{Yichen Yang}
\author[1]{Olivier Pierron}
\author[1]{Josh Kacher}
\author[2]{David Rowenhorst\corref{cor1}}
\ead[url]{david.j.rowenhorst.civ@us.navy.mil}

\affiliation[1]{organization={Georgia Institute of Technology},
addressline={School of Materials Science and Engineering},
postcode={30332},
city={Atlanta, GA},
country={USA}}

\affiliation[2]{organization={US Naval Research Laboratory},
addressline={Materials Science and Technology Division},
postcode={20375},
city={Washington, DC},
country={USA}}

\cortext[cor1]{Corresponding author}



\begin{abstract}
4D-STEM-based orientation and phase mapping has enabled rapid microstructure quantification that can be directly combined with standard TEM- and STEM-based imaging modes. Typically, orientation mapping is coupled with beam precession (i.e. precession electron diffraction) to achieve high indexing rates, adding to the cost and often decreasing the spatial resolution of the approach.  This paper introduces a new post processing approach modeled after the non-local pattern averaging and reindexing algorithm developed for the electron backscatter diffraction community, wherein post-collection, patterns are averaged using a distance similarity parameter. Results from Ni and Au thin films show that indexing rates can be significantly improved using this post-processing technique due to improved signal-to-noise ratios in the diffraction patterns. Interestingly, the highest indexing rates are achieved in samples heavily damaged via ion irradiation, suggesting that averaging over curved lattices further improves indexing rates.
\end{abstract}

\end{frontmatter}
\section{Introduction}

High-speed electron detectors have ushered in a new era of data-rich electron microscopy, spurring the incorporation of data science tools into TEM analysis~\cite{muller2012scanning,ercius20204d,spurgeon2021towards,nord2020fast}. One of the areas in which this is most apparent is in the development of scanning nanobeam diffraction techniques, commonly referred to as 4D-STEM~\cite{ophus2019four}. In this approach, a grid of nanobeam diffraction patterns is collected over a defined area and stored for offline analysis. Virtual bright and dark field images are created, elastic strain gradients are calculated, and ptychography approaches are applied.

Pattern indexing to resolve the local phase and orientation state has been developed and is widely available in commercial and open source packages~\cite{cautaerts2022free,rauch2010automated,ophus2019four}. This approach is similar to EBSD in that it indexes diffraction patterns by comparing to a library (triplet indexing for EBSD and template matching for 4D-STEM) to determine the crystal orientation and structure associated with the electron beam/material interaction point. However, orientation/phase mapping using 4D-STEM patterns presents unique difficulties not associated with EBSD pattern indexing. Specifically, the pattern captures a much smaller fraction of diffraction space, and the templates used for template matching do not contain the dynamical diffraction information that is contained in most experimental diffraction patterns. For this reason, a number of hardware and software approaches have been developed to facilitate accurate microstructure characterization.

On the hardware side, precession electron diffraction (PED) is routinely coupled with orientation/phase mapping~\cite{rauch2010automated,midgley2015precession}. The beam precession increases the fraction of reciprocal space probed and averages out dynamical diffraction effects. In addition, patterned apertures have been developed to counteract the influence of intensity variations within the diffraction disks~\cite{zeltmann2019improved,guzzinati2019electron}. While effective, these hardware options can add significant cost to instrumentation, are not available in many facilities, and, in the case of PED, can increase scan acquisition times while decreasing the spatial resolution.

On the software side, a number of algorithms have been developed to improve indexing reliability and more accurately resolve local microstructure. Several groups have developed clustering algorithms for 4D-STEM data with the aim of reducing the dimensionality (and space requirements) of the diffraction data while still accurately capturing local microstructural features~\cite{kimoto2024unsupervised,bruefach2023robust,shi2022uncovering,bergh2020nanocrystal}. These algorithms have been demonstrated to accurately capture local features, but are not designed to improve reliability in the identification of the crystal structure or phase of those features beyond what standard template matching algorithms are able to accomplish. Lee et al. recently developed an approach using unsupervised clustering and pattern averaging to improve orientation and phase indexing rates while also substantially reducing the volume of needed data~\cite{lee2026unsupervised}.

In this paper, we present a new algorithm-based approach for non-local smoothing of 4D-STEM diffraction patterns that can dramatically increase orientation indexing rates while still resolving local microstructural variations. This approach is modified from the NLPAR approach developed for the EBSD community~\cite{brewick2019nlpar}, which was developed to improve EBSD-based orientation and phase indexing rates while preserving edges and sharp features. We briefly presented a proof-of-concept of this work in~\cite{Rowenhorst2025p589}, and here we more fully explain the methodology and results of this new algorithm  developed for 4D-STEM pattern processing, which we refer to as NLSTEM.

\section{NLSTEM algorithm}

The concepts and implementation of the NLSTEM algorithm are largely similar to the original implementation found in Brewick et al.~\cite{brewick2019nlpar}, with some important adaptations specific to working with 4D-STEM datasets. This section will briefly summarize the algorithm presented by Brewick et al., followed by a more detailed description of the differences between the NLPAR and NLSTEM algorithms. 

 The original NLPAR algorithm was based off the non-local means algorithm developed by Buades et al.~\cite{Buades2005p60,Buades2005p490} and was further adapted for hyper-spectral images, namely EBSD patterns, and now 4D-STEM patterns. Broadly, the NLSTEM algorithm calculates the similarity between a pattern of interest, $\mathbf{p}_i$, and other diffraction patterns in a dataset, generates a weighting function based off of the similarity, and uses the weighting function to average the diffraction patterns together. Each diffraction pattern is considered a 1-D vector of the image intensities, $\mathbf{p}$, with length $N_p$. The goal of the algorithm is to find a reasonably large set of similar diffraction patterns that share much of the same information as $\mathbf{p}_i$ while excluding those that have dissimilar information. One could evaluate the similarity between all the patterns within the dataset, but it is much more practical to look in a window that is likely to have similar patterns. Here a user defined search window, $\mathcal{W}$, is selected around $\mathbf{p}_i$. 
For each pattern within $\mathcal{W}$, the $L2$ norm is calculated between $\mathbf{p}_i$ and each of the other patterns contained within $\mathcal{W}$, notated here as $\mathbf{p}_j$. As Buedes et al.\ showed~\cite{Buades2005p490}, for a constant noise ratio added to the original pattern signals, the $L2$ norm is normally distributed, with the expected value equal to the variance of the added noise.  Brewick et al.\ showed that this could be extended for the instance where the noise pattern is not the same for the two patterns.  This distance metric is then used to calculate an exponentially decaying normalized weighting factor, $w(i,j)$ that is used to average the patterns within $\mathcal{W}$: 

\begin{align}
{d}(\mathbf{p}_i,\mathbf{p}_j) &= \frac{\sum_{k}^{N_{p}} \left(\mathbf{p}_i[k]\! - \! \mathbf{p}_j[k]\right)^{2}\!-\!N_{p} \left(\sigma^2_i\! +\! \sigma^2_j\right)} {\sqrt{2N_{p}}\left(\sigma^2_i\! +\! \sigma^2_j\right)} \label{eqn:distance}
\\
w(i,j) &= \frac{1}{Z(i)}\exp\left(-\frac{\max\left({d}\left(\mathbf{p}_i,\mathbf{p}_j\right),  0\right)}{\lambda^2}\right) \label{eqn:weights_sub}
\\
Z(i) &= \sum_{j}^\mathcal{W} \exp\left(-\frac{\max\left({d}\left(\mathbf{p}_i,\mathbf{p}_j\right),  0\right)}{\lambda^2}\right)
\end{align}
where $N_{q}$ is the number of pixels in the pattern image, $d$ is the normalized distance between the pattern images, $Z$ is a normalization factor for the weights, $\sigma$ provides an estimate of the amount of noise associated with each pattern, and $\lambda$ is a user chosen parameter that controls the decay of the weights. Careful examination of Eq.~\ref{eqn:weights_sub} shows that the weighting function can be interpreted as expressing the probability that the two patterns are from the same sampling.  This then explains the insertion of the maximum function for normalized distances below zero, as it would be highly likely those patterns are indeed from the same sampling and thus should be averaged at the same weight as $\mathbf{q}_i$. 
The resultant NLSTEM pattern, $\mathbf{p}_i'$, is given by the weighted average of all the patterns in the window: 
\begin{equation}
\begin{aligned}
\mathbf{p}'_i[k] &= \sum_j^{\mathcal{W}} w(i,j)\mathbf{p}_j[k] \\
\end{aligned}
\end{equation}\label{eqn:weighted average} 
This process is evaluated for every pattern within the 4D-STEM scan. On a modern multi-processor workstation this only requires on the order of minutes to complete for most datasets.

There are two important differences when considering 4D-STEM diffraction patterns compared to EBSD patterns. First, the noise distribution is significantly different in the two patterns. In EBSD, the noise associated with each pattern is modeled as normally-distributed random noise added to the true signal.  In the case of EBSD patterns, this is generally considered a valid assumption as the noise is typically dominated by a mixture of detector noise, read-out noise, and, since the dynamic range of the EBSD patterns is relatively low, Poisson noise that closely resembles normally distributed noise for all intensities. However, in the case of 4D-STEM diffraction patterns, there is a much higher dynamic range, and with most detectors, comparably low detector noise. This is especially true when collecting patterns using a direct detection camera.  Therefore, the noise is dominated by Poisson sampling noise, where the magnitude of the noise is proportional to the square-root of the intensity.  Thus, while the signal to noise ratio decreases with increasing intensity, the absolute magnitude of the noise, and thus the relative difference between two images, will be higher for higher intensity regions in the pattern (i.e. diffraction spots). To standardize the noise across all intensities, the patterns are scaled by the square root of the original intensity:
\begin{equation}
   \mathbf{q} = \sqrt{\mathbf{p}} 
\end{equation}
with $\mathbf{q}_j$ replacing $\mathbf{p}_j$ in the previous equations. After calculating the weighted average, the intensities are scaled back to the original dynamic range according to:
\begin{equation}
\begin{aligned}
\mathbf{p}_i' &= \left(\mathbf{q}_i'\right)^{2}
\end{aligned}
\end{equation}\label{eqn:rescaled} 

It should be noted that this is only an approximation of the noise associated with the images, and should not be considered a full analysis of all the sources of noise within the diffraction patterns. In our testing, the fine details of the noise model did not have a large impact on the performance of the NLSTEM algorithm.  

The second important difference between EBSD patterns and 4D-STEM diffraction patterns in the implementation of NLSTEM is the difference in intensity variations. 4D-STEM diffraction patterns are dominated by a few discrete, high intensity diffraction spots that are near the detector saturation level. In addition, the direct beam spot is a dominant and near-constant feature in all diffraction patterns, which can strongly affect the weighting algorithm. For portions of the pattern that are so bright that the detector is saturated, such as the direct beam in the 4D-STEM diffraction patterns, the measured noise is zero. Any comparison metric that then includes these saturated regions would produce an artificially high degree of similarity (equivalent to a low value of $d$), perhaps even suppressing other significant differences in the rest of the pattern and thereby erroneously combining non-similar patterns.  This is handled within NLSTEM in two different ways.  First, a mask is created for the direct beam (either by examining where the brightness beam is consistently within the patterns, or by examining the metadata for the STEM patterns if the direct beam location is included), where points within the mask are not included in the distance metric, Eq.~\ref{eqn:distance}. Secondly, the direct beam may not be the only saturated spot within the pattern image. Thus, for each pattern pair, the distance calculation excludes the pixel locations where either of the pattern intensities are within $1\,\%$ of the maximum intensity measured within $\mathcal{W}$.  This $1\,\%$ rule might lead to excluding non-saturated points for a set of patterns that have no saturation, but this would have a very minor impact as it only eliminates a small amount of the total pixel comparison data between the two patterns.  It should also be noted that this means that $N_q$ is no longer a constant for all pattern pairs.    

One application note should be mentioned here. Some manufactures store the intensity in the diffraction patterns as a floating point number with both positive and negative values, and will separately store the intensity scaling factors in separate metadata.  Here the data is maintained as floating point numbers, but to avoid a numerical error when scaling the intensity by the square-root,  for each grouping of patterns the minimum value is subtracted off the entire group for the NLSTEM algorithm, then after processing, the NLSTEM patterns are restored to their original range by adding the pre-processed group minimum back onto the intensities. 

The user is left with only a few adjustable parameters, namely, $\lambda$, $\sigma$ and the size of $\mathcal{W}$.  For $\sigma$, Brewick et al.\ showed that by making the reasonable assumption that nearest-neighbor patterns have a high probability of being nearly identical to the pattern of interest other than noise, one can easily generate reasonable estimates for $\sigma$ for each pattern. They also provided a method using this same assumption to provide estimates for $\lambda$, which,  with the above normalization of the distances in Eq.~\ref{eqn:distance}, it should be $\approx 1$.  In this work, we use the optimized value of $\lambda = 1.07$ for all datasets to provide consistency across the experiments. The choice of window size ends up being a heuristic that should consider how likely it is to find similar patterns within the window, the allowable calculation time, and the amount of memory available for the NLSTEM calculation.  Here again to maintain consistency, we chose a constant square sliding window with a search radius, $\mathcal{SR}= 4$, which means each pattern is being compared to a total of $\left(2\mathcal{SR}+1\right)^2-1 = 80$ neighbors.

\section{Methods}

Two different material sets were used for validation of the NLSTEM algorithm: nanocrystalline Ni and ultrafine  grained Au irradiated to different damage levels. 4D-STEM datasets were collected in a ThermoFisher Tecnai F30 operated at 300\,kV in STEM mode. 4D-STEM datasets were acquired with a convergence semi-angle of 0.7~mrad at a camera length of 150~mm using a Gatan Metro~300 pixelated detector. A spot size of 9 with dwell times of 3~ms per pixel was used for pattern acquisition.

Freestanding nanocrystalline Ni and ultra-fine grained Au films (thickness 100~nm) were deposited by electron-beam evaporation at a base pressure of $\sim\!10^{-6}$~Torr and a deposition rate of 1~\AA/s at room temperature. Ex~situ ion irradiation for Au films was carried out with a 2.8~MeV $\mathrm{Au}^{4+}$ beam to a fluence of $5.5\times10^{13}$~ions/cm$^{2}$. Samples were irradiated to damage levels of 0 (as-fabricated), 1, and 5 displacements per atom (dpa), where dpa refers to the average displacement events each atom undergoes (see~\cite{stangebye2023direct,DAZALLANOS2025149256} for additional details on the sample fabrication, irradiation procedures, and damage estimates).

After collecting the datasets, the diffraction patterns were indexed to retrieve the local crystallographic orientation using Gatan's STEMx orientation mapping software. This software uses a template matching algorithm similar to what is used in open source packages such as py4DSTEM~\cite{savitzky2021py4dstem}. Patterns were indexed before and after applying NLSTEM. For one comparison scan, each pattern was averaged with its four nearest neighbors rather than using a non-local smoothing approach. This is equivalent to the neighbor pattern averaging and reindexing (NPAR) approach developed in the EBSD community~\cite{wright2015introduction} and functionally is the same as setting $\lambda$ equal to $\infty$ in the NLSTEM algorithm. The indexing rate was determined by filtering out all datapoints with a confidence index below 0.4, where the confidence is based on the correlation factor between the template and the collected diffraction pattern. This value was chosen based off of past experiments showing 0.4 to be a reasonable cut off for reliable indexing. Inverse pole figure (IPF) maps were constructed using OIM Analysis.

\begin{figure*}[tb]
    \centering
    \includegraphics[width=0.9\linewidth]{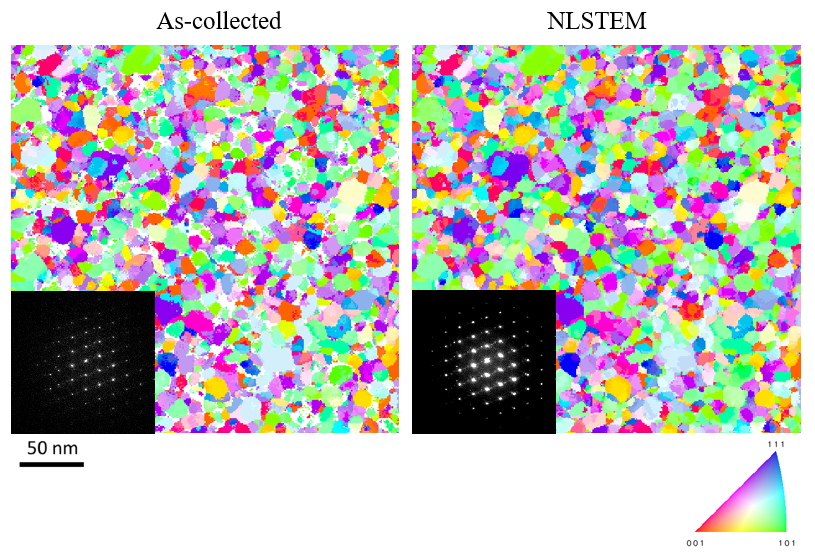}
    \caption{4D-STEM orientation maps of a nanocrystalline Ni thin film obtained from as-collected data and after NLSTEM processing. (a)~As-collected IPF orientation map. Inset in~(a) is a representative raw convergent-beam electron diffraction pattern, illustrating the faint, noisy Bragg discs caused by the nanoscale grain mixture. (b)~Corresponding IPF orientation map after NLSTEM processing of the same dataset, yielding a nearly fully indexed map. The enhanced diffraction pattern quality is shown in inset in~(b)}
    \label{fig:Ni}
\end{figure*}

\section{Results}

\subsection{Nanocrystalline Ni}

We first evaluate NLSTEM on a nanocrystalline Ni thin film, a regime where orientation mapping is hindered by pattern mixing. With thickness $t\approx100$~nm and mean grain size $\bar D\approx17$~nm ($t/\bar D\sim 6$), a single probe position often intersects multiple grains. Figure~\ref{fig:Ni} contrasts the as-collected and NLSTEM-processed IPF maps. The raw dataset reaches only $90.6\,\%$ indexing, with unindexed patches near grain boundaries and triple junctions. After NLSTEM, the indexed fraction rises to $>\!99.9\,\%$, yielding contiguous grain interiors while retaining sharp grain-boundary contrast. A representative diffraction pattern is included from each scan, visually demonstrating the improved signal to noise in NLSTEM processed pattern (more detailed analysis of individual diffraction patterns is included in the analysis of the Au films). Extra spots in the diffraction patterns reflect multi-grain through thickness nature of the thin films as the diffracted electrons originate from more than one crystallographic orientation.

\begin{figure*}[tbh]
    \centering
    \includegraphics[width=0.9\linewidth]{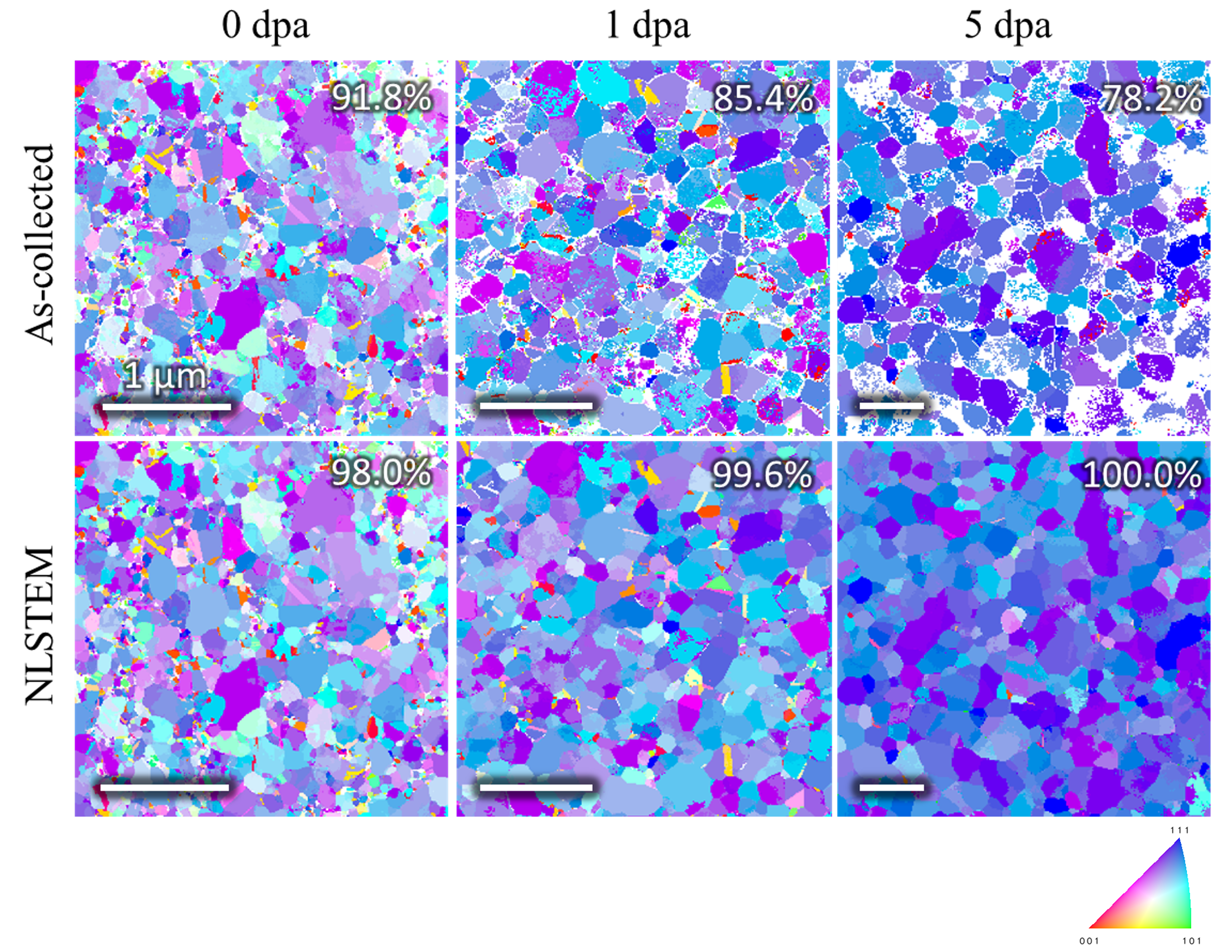}
    \caption{Effect of NLSTEM processing on 4D-STEM crystal orientation maps of irradiated Au thin films at 0, 1, and 5~dpa. Top row: as-collected indexing results; bottom row: after NLSTEM processing and re-indexing. The precent of reliably indexed data points is included with each map.}
    \label{fig:Au}
\end{figure*}

\subsection{Irradiated Au}
To assess how irradiation-induced lattice damage affects indexing, 4D-STEM scans were acquired from Au thin films at 0~dpa (as-deposited), 1~dpa, and 5~dpa (Figure~\ref{fig:Au}). In the as-collected maps (top row), indexed coverage decreases systematically with dose: at 0~dpa, unindexed pixels are mainly confined to grain boundaries, whereas at 5~dpa large unindexed (white) regions extend into grain interiors. This degradation is consistent with defect structures (e.g., stacking-fault tetrahedra and small dislocation loops) generating local strain fields and diffuse scattering that reduces Bragg-disk contrast and undermines template matching (see~\cite{DAZALLANOS2025149256} for more detailed TEM analysis of the defect structures). After NLSTEM processing and re-indexing (bottom row), grain interiors become contiguous and boundaries are more sharply delineated at all doses, producing a pronounced recovery in indexing coverage.

Surprisingly, the indexing rate at 5~dpa after NLSTEM processing is higher than the indexing rate in the as-deposited sample after NLSTEM processing. There are two potential explanations for this increased indexing rate. First, ion irradiation coarsens the grains, reducing the fraction of boundary/mixed pixels. Within the non-local search window this raises the likelihood of same-orientation neighbors, strengthening similarity weights and suppressing speckle without shifting Bragg-disk maxima. This grain coarsening largely eliminates the smallest grains from the matrix, which are the most likely to lead to mixed diffraction patterns. Secondly, work has shown that ion irradiation can lead to internal grain curvature~\cite{yu2021new}. Defect-mediated lattice curvature between adjacent probe positions could act as a small-angle angular integration. Similarly to PED processing, the similarity-weighted averaging damps dynamical oscillations while preserving peak positions.

 \begin{figure*}[tb!]
    \centering
    \includegraphics[width=0.9\linewidth]{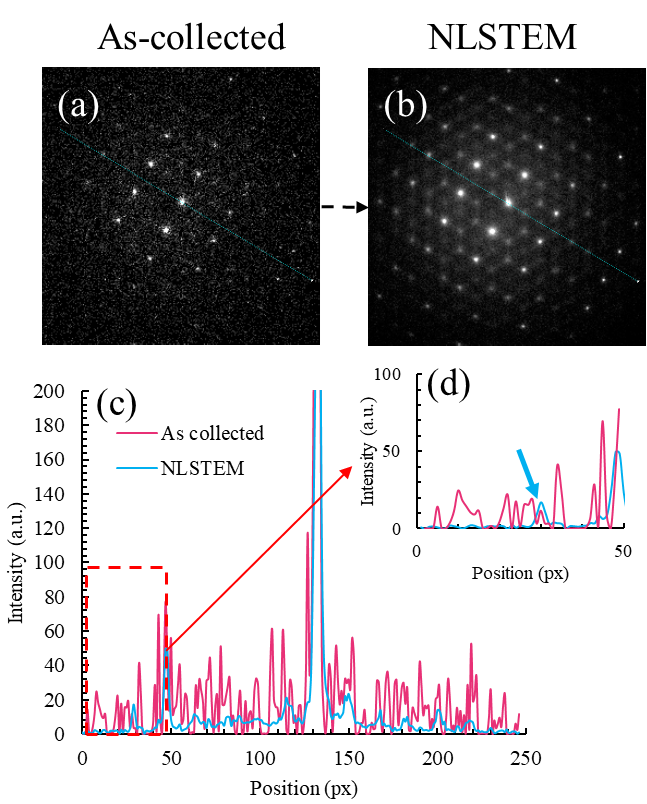}
    \caption{Effect of NLSTEM processing on a representative diffraction pattern from 5~dpa irradiated Au. (a)~Diffraction pattern prior to NLSTEM processing. (b)~The same diffraction pattern after NLSTEM processing. (c)~Intensity profile taken along the blue lines in~(a) and~(b) showing the increased signal to noise ratio after NLSTEM processing. (d)~Segment of the intensity profile shown in boxed region in~(c) highlighting minor diffraction peaks that can be identified after NLSTEM processing. A blue arrow is used to indicated one example of a faint peak only identifiable in the post-processed pattern.}
    \label{fig:Au Pattern}
\end{figure*}

\section{Discussion}
\subsection{Mechanisms for improved indexing}

The above results demonstrate clear improvements in indexing rates in ultrafine grained and nanocrystalline materials after NLSTEM processing. Implementation of this algorithm is straightforward and can be used in conjunction with any 4D-STEM-based orientation mapping software (Gatan STEMx, py4DSTEM...). In addition, while not shown here, it is expected that implementing NLSTEM processing would also improve indexing rates in PED data or when using patterned apertures. The improvement in indexing is due to the improved signal to noise ratio of the diffraction patterns, demonstrated more clearly in  Figure~\ref{fig:Au Pattern}. Here, a detailed comparison of a representative diffraction pattern before and after NLSTEM processing taken from the 5~dpa irradiated Au sample is shown. As is visually apparent, the noise level, in the form of white speckles, decreases significantly after NLSTEM processing. In addition, diffraction peaks that are not visible before processing, including faint peaks and high-order reflections, become easily visible after processing. This effect is more apparent in the line intensity profiles shown in Fig.~\ref{fig:Au Pattern}c-d. The high noise level before processing drowns out many of the diffraction peaks, such as the one indicated by an arrow in Fig.~\ref{fig:Au Pattern}d. The increased prevalence of diffraction peaks and decreased background noise levels directly improve the success rate in template matching algorithms and decrease the propensity for false matches~\cite{wang2023improving,zhao2023reference}. The improvement in indexing rates is most noticeable in the ultrafine grained Au where the larger grain size decreases the likelihood of pattern mixing from multiple through thickness grains and increases the population of nearby similar diffraction patterns.

The results from the 5~dpa Au sample suggest that a secondary mechanism for improved pattern indexing is through an inverted PED mechanism. When averaging diffraction patterns via NLSTEM-processing in ductile materials such as metal thin films, there will almost invariably be small crystallographic rotations in neighboring points, either because of geometrically necessary dislocations or due to bending in the thin film. As a result, the NLSTEM-processed diffraction patterns are averaged over a small range of crystallographic orientations, similar to pattern acquisition during PED. This reduces dynamic effects in the overall diffraction pattern, resulting in a quasi-kinematic pattern~\cite{midgley2015precession}. The introduction of nanoscale irradiation defects is expected to further increase local lattice strains and rotations~\cite{yu2021new}, magnifying this inverted PED mechanism and leading to improved indexing rates in ion-irradiated Au films in comparison to as-deposited thin films.

\begin{figure*}[bt!]
    \centering
    \includegraphics[width=0.9\linewidth]{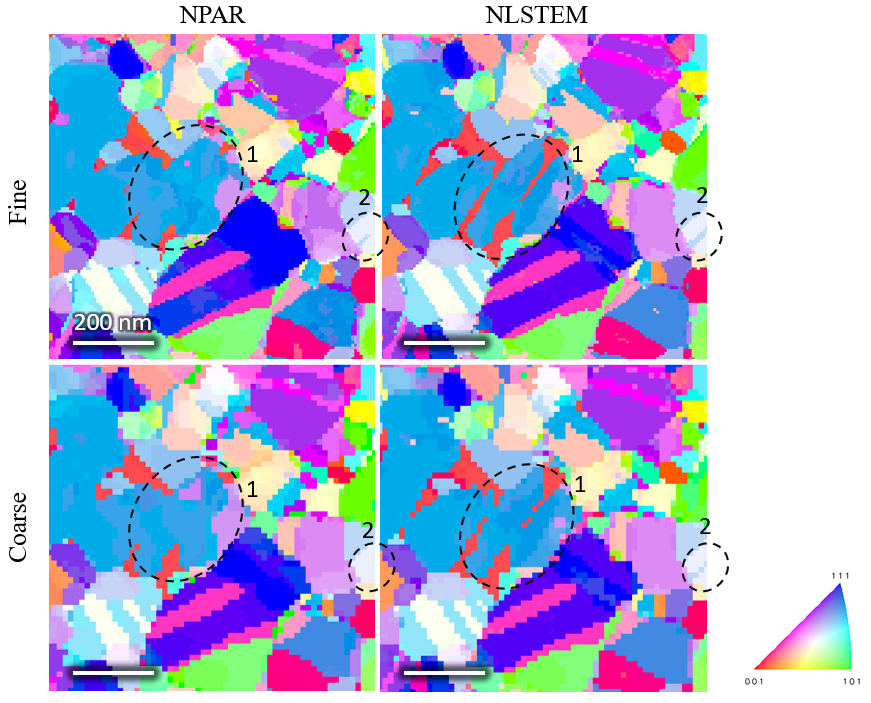}
    \caption{IPF maps of the same area in an annealed Ni thin film indexed using NPAR and NLSTEM under two pattern-collection step sizes: 10~nm (fine) and 20~nm (coarse). Dashed circles mark thin twin lamellae with thicknesses of approximately $\sim\!10$~nm~(1) and $\sim\!20$~nm~(2).
    }
    \label{fig:NPAR}
\end{figure*}

\subsection{Comparison between NLSTEM and nearest neighbor averaging}

NLPAR was originally developed in the EBSD community to overcome the loss in edge and sharp features associated with NPAR algorithms~\cite{brewick2019nlpar}. While NPAR is computationally less expensive, it indiscriminately averages nearest neighbor patterns, leading to small-scale features being averaged out in the analysis, and simultaneously limiting the size of the averaging window.  As a comparison between the algorithms, Figure~\ref{fig:NPAR} depicts the same region of an annealed Ni thin film indexed using NPAR and NLSTEM using two different pattern-collection step sizes: a fine step size of 10~nm (top row) and a coarse step size of 20~nm (bottom row). Two twin lamellae contained within the scans are highlighted: a thinner twin in circle~1 (width of $\sim\!10$~nm) and a thicker twin in circle~2 (width of $\sim\!20$~nm). At the 5~nm step size, NPAR resolves the $\sim\!20$~nm twin (circle~2) as a continuous lamella with a consistent IPF color contrast relative to the parent grain, but only the thicker regions of the $\sim\!10$~nm twin (circle~1) are fully resolved. In contrast, NLSTEM resolves both twins (circles~1 and~2) as continuous lamellae with clear, coherent contrast across their full lengths.

When the step size is increased to 10~nm, the NPAR result further degrades. The $\sim\!20$~nm twin in circle~2 is no longer visible within the grain and the $\sim\!10$~nm twin in circle~1 remains unresolved. NLSTEM at 10~nm step size still retains both twins as identifiable lamellae, with the bands remaining continuous despite the coarser sampling. This demonstrates that NLSTEM is able to capture small-scale features (features on the order of a step size) that would be lost using more a straightforward NPAR approach.

\section{Conclusion}

We introduce the NLSTEM algorithm for improving template matching-based orientation mapping in 4D-STEM. The algorithm, based off of the NLPAR algorithm from the EBSD community, shows marked improvements in indexing rates, with improvements most evident in ultrafine grained materials. Interestingly, lattice damage is shown to improve indexing rates after NLSTEM processing, suggesting a inverted PED effect. By comparing the performance of NPAR and NLSTEM algorithms, it was shown that NLSTEM is able to improve indexing rates while also retaining features with dimensions on the order of the scan step size that would otherwise be lost with simple neighbor pattern averaging. The source code is made freely available at \url{https://github.com/USNavalResearchLaboratory/PyEBSDIndex} and is compatible with any 4D-STEM template-matching software.

\section{Acknowledgment}
JK, OP, and YY gratefully acknowledge support by the U.S. Department of Energy (DOE), Office of Science, Basic Energy Sciences (BES) Materials Science and Engineering (MSE) Division under Award \#DE-SC0018960.  DJR gratefully acknowledges the support of the US Naval Research Laboratory Base Program.

\bibliographystyle{elsarticle-num} 
\bibliography{Refs}
\end{document}